\documentstyle[emulateapj,amsfonts,psfig,natbib,amsmath]{article}

\def\swift{{\it Swift}}
\def\grb{GRB\,051221a}

\def\cit{1}
\def\ociw{2}
\def\prince{3}
\def\hubble{4}
\def\nrao{5}
\def\uh{6}
\def\anu{7}
\def\psu{8}
\def\srl{9}
\def\tapir{10}
\def\gem{11}

\begin{document}

\title{\large The Afterglow, Energetics and Host Galaxy of the Short-Hard 
Gamma-Ray Burst 051221a}

\author{
A.~M.~Soderberg\altaffilmark{\cit},
E.~Berger\altaffilmark{\ociw,}\altaffilmark{\prince,}\altaffilmark{\hubble},
M.~Kasliwal\altaffilmark{\cit},
D.~A.~Frail\altaffilmark{\nrao},
P.~A.~Price\altaffilmark{\uh},
B.~P.~Schmidt\altaffilmark{\anu},
S.~R.~Kulkarni\altaffilmark{\cit},
D.~B.~Fox\altaffilmark{\psu},
S.~B.~Cenko\altaffilmark{\srl},
A.~Gal-Yam\altaffilmark{\cit,}\altaffilmark{\hubble},
E.~Nakar\altaffilmark{\tapir},
and K.~C.~Roth\altaffilmark{\gem}
}

\altaffiltext{\cit}{Division of Physics, Mathematics and Astronomy,
105-24, California Institute of Technology, Pasadena, CA 91125}

\altaffiltext{\ociw}{Observatories of the Carnegie Institution
of Washington, 813 Santa Barbara Street, Pasadena, CA 91101}
 
\altaffiltext{\prince}{Princeton University Observatory,
Peyton Hall, Ivy Lane, Princeton, NJ 08544}
 
\altaffiltext{\hubble}{Hubble Fellow}

\altaffiltext{\nrao}{National Radio Astronomy Observatory, Socorro,
NM 87801}

\altaffiltext{\uh}{Institute for Astronomy, University of Hawaii, 
2680 Woodlawn Drive, Honolulu, HI 96822}

\altaffiltext{\anu}{Research School of Astronomy and Astrophysics, 
Australian National University, Mt Stromlo Observatory, via Cotter Rd,
Weston Creek, ACT 2611, Australia}

\altaffiltext{\psu}{Department of Astronomy and Astrophysics, 
Pennsylvania State University, 525 Davey Laboratory, University 
Park, PA 16802}

\altaffiltext{\srl}{Space Radiation Laboratory, MS 220-47, California 
Institute of Technology, Pasadena, CA 91125}

\altaffiltext{\tapir}{Theoretical Astrophysics, 130-33 
California Institute of Technology, Pasadena, CA 91125}

\altaffiltext{\gem}{Gemini Observatory, 670 N. Aohoku Place Hilo, HI 
96720}

\begin{abstract}
We present detailed optical, X-ray and radio observations of the
bright afterglow of the short gamma-ray burst 051221a obtained with
Gemini, {\it Swift}/XRT, and the Very Large Array, as well as
optical spectra from which we measure the redshift of the burst,
$z=0.5464$.  At this redshift the isotropic-equivalent prompt energy
release was about $1.5\times 10^{51}$ erg, and using the standard
afterglow synchrotron model we find that the blastwave kinetic energy
is similar, $E_{K,\rm iso}\approx 8.4\times 10^{51}$ erg.  An
observed jet break at $t\approx 5$ days indicates that the opening
angle is $\theta_j\approx 7^\circ$ and the total beaming-corrected
energy is therefore $\approx 2.5\times 10^{49}$ erg, comparable to the
values inferred for previous short GRBs.  We further show that the
burst experienced an episode of energy injection by a factor of 3.4
between $t=1.4$ and 3.4 hours, which was accompanied by reverse shock
emission in the radio band.  This result provides continued evidence
that the central engines of short GRBs may be active significantly
longer than the duration of the burst and/or produce a wide range of
Lorentz factors.  Finally, we show that the host galaxy of
GRB\,051221a is actively forming stars at a rate of about 1.6
M$_\odot$ yr$^{-1}$, but at the same time exhibits evidence for an
appreciable population of old stars ($\sim 1$ Gyr) and near solar
metallicity.  These properties are intermediate between those of long
GRB hosts and those of previous short bursts suggesting that the
progenitor lifetimes may have a large spread. The lack of bright
supernova emission and the low circumburst density ($n\sim 10^{-3}$
cm$^{-3}$), however, continue to support the idea that short bursts
are not related to the death of massive stars and are instead
consistent with a compact object merger.  Given that the total energy
release is a factor of $\sim 10$ larger than the predicted yield for a
neutrino annihilation mechanism, this suggests that
magnetohydrodynamic processes may be required to power the burst.
\end{abstract}

\keywords{gamma-ray bursts: specific (GRB\,051221a)}

\section{Introduction}
\label{sec:intro}
Following over a decade of speculation about the nature of the
short-duration hard-spectrum gamma-ray bursts (GRBs), the recent
discovery of afterglow emission from these bursts provided the first
physical constraints on possible progenitor models.  We now know that
short GRBs occur at cosmological distances, $z\gtrsim 0.1$, have
energy releases of about $10^{49}$ erg when corrected for relatively
wide opening angles of $\sim 10^\circ$, are associated with both
star-forming and elliptical galaxies, and are not associated with
bright supernovae
\citep{bcb+05,bpc+05,bpp+05,ffp+05,gso+05,hsg+05,hwf+05,vlr+05}.  
These properties indicate that short GRBs are associated with an old
stellar population, and are broadly consistent with expectations of
the popular neutron star (NS) and/or black hole (BH) coalescence
models (e.g., \citealt{elp+89,npp92,kc96,rr02,ajm05}).

On the other hand, some of the afterglow observations, in particular
the detection of soft X-ray tails and X-ray flares on a timescale
$\gtrsim 10^2$ s \citep{bcb+05,vlr+05}, cannot be easily explained in
the coalescence scenario.  These observations raise the possibility
that short GRBs may have different or multiple progenitors systems
(e.g., accretion-induced collapse of a neutron star; \citealt{mrz05}),
or that some key physical processes in the energy extraction mechanism
or accretion disk hydrodynamics remain unaccounted for in the simple
models \citep{paz06}.  Secondary indicators such as the redshift and
luminosity distributions have also been used to argue against the
coalescence model (\citealt{ngf05}, but see \citealt{gp05}), but these
rely on a very limited sample of events ($\lesssim 4$ with reasonably
robust redshifts) and strong constraints are likely premature.

In the absence of direct evidence for or against the coalescence model
from the detection of gravitational waves (e.g., \citealt{ct02}), or a
mildly relativistic outflow \citep{lp98,k05}, progress in our
understanding of the nature and diversity of the progenitor systems
and the underlying physics requires a much larger statistical sample
than the current data set of two short GRBs with secure redshifts
(050709 and 050724) and two with putative redshifts (050509b and
050813).  In addition, only GRBs 050709 and 050724 have broad-band
afterglow coverage which allowed a determination of the energy release
and circumburst density, and provided evidence for ejecta collimation
\citep{bpc+05,ffp+05,p05}.  With a larger sample of well-studied 
short GRBs we can begin to address the properties of the progenitor
systems and the GRB mechanism through a census of the energy release,
the jet opening angles, the density of the circumburst environment,
and the relative fraction of short GRBs in star-forming and elliptical
galaxies.

In this vein, we present in this paper detailed optical, X-ray and
radio observations of the bright \grb, which we localize to a
star-forming galaxy at $z=0.5464$.  We show from the evolution of the
various light curves and from modeling of the broad-band afterglow
data that the beaming-corrected energy release of this burst is 
comparable to those of previous short-hard bursts.  In
addition, we find that the circumburst density is low indicating that
the progenitor system is unlikely to reside in a star forming region.
Finally, we show that the GRB experienced an episode of significant
energy injection, which was accompanied by reverse shock emission in
the radio band.  This, along with an energy release
exceeding $10^{49}$ erg, suggests that the energy extraction mechanism
leads to a complex ejecta structure and is possibly driven by MHD
processes rather than neutrino anti-neutrino annihilation.

\section{Observations}
\label{sec:obs}
GRB\,051221a was discovered by the {\it Swift} Burst Alert Telescope
(BAT) on 2005 December 21.0773 UT \citep{pbb+05} and localized to a
$0.8'$ radius error circle centered on $\alpha=21^{\rm h}
54^{\rm m} 50.7^{\rm s}$, $\delta=16^{\circ} 53' 31.9''$ (J2000;
\citealt{cbb+05}).  The burst was also observed by the Suzaku Wideband
All-sky Monitor (WAM; \citealt{eta+05}) and by Konus-Wind \citep{gam+05}.
The gamma-ray light-curve is characterized by an initial hard pulse
(FWHM $\sim 0.25$ s) composed of several separate pulses (FWHM $\sim
15$ ms) followed by a softer emission component lasting $\sim 1$ s
\citep{nsb+05,eta+05,gam+05}.  The total duration of the burst is
$T_{90}\approx 1.4\pm 0.2$ sec with a peak energy, $E_p\approx 400\pm
80$ keV \citep{cbb+05,gam+05}. The short duration and high $E_p$
indicate that \grb\ is a classical short burst.  The Konus-Wind
gamma-ray fluence is $F_{\gamma}=3.2^{+0.1}_{-1.7}\times 10^{-6}~\rm
erg~cm^{-2}$ (20 keV -- 2 MeV; \citealt{gam+05}).

Starting at $t\approx 92$ sec the X-ray Telescope (XRT) onboard {\it
  Swift} detected a fading source at $\alpha=21^{\rm h} 54^{\rm m}
48.71^{\rm s}$, $\delta=16^{\circ} 53' 28.2''$ (J2000) with a $90\%$
containmemt error radius of $3.5''$ \citep{bgc+06}.  A near-infrared
(NIR) object was identified within the XRT error circle although it
showed no evidence for fading between $t\approx 1.19$ and 1.38 hr
after the burst \citep{b05a,b05b}.

\subsection{Optical Observations}
\label{sec:optical}

\begin{figure}
\vspace{-1.75cm}
\plotone{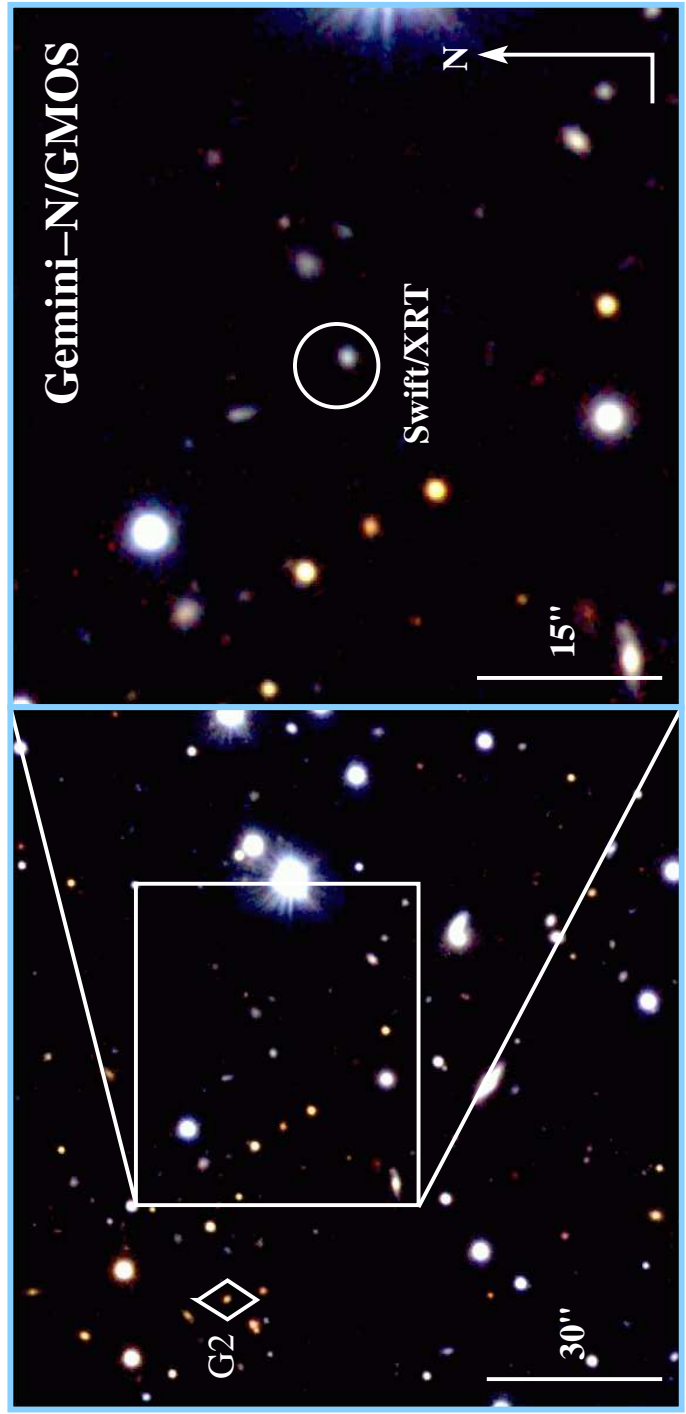}
\vspace{-1.75cm}
\caption{Color composite image of the field of \grb\ as observed with
  Gemini-N/GMOS.  The image is composed of $i'$- and $z'$-band data
  from 2005 Dec 27.2 UT and $r'$-band data from 2005 Dec 30.2 UT.  The
  host galaxy of \grb\ is the only source detected inside the XRT
  error circle and is clearly blue as expected for a star-forming
  galaxy.  We note that our spectrum of \grb\ also included galaxy G2
  (diamond, left panel).  This galaxy exhibits a strong Balmer/4000\AA\
  break in addition to \ion{Ca}{2} H\&K absorption at $z=0.849$.  The
  presence of several nearby galaxies with a similar color and
  brightness may point to a background cluster unassociated with the
  burst.}
\label{fig:color_field}
\end{figure}

We began observations of \grb\ in the $r'$-band using the Gemini
Multi-Object Spectrograph (GMOS) on the Gemini-north 8-m telescope
about 2.8 hr after the burst and detected a bright optical source
coincident with the NIR object (Figure~\ref{fig:color_field}).  A
comparison of the first and last exposures in the observation (227 and
271 min after the burst) revealed that the source has faded indicating
that it is the optical afterglow of \grb.  Astrometry was performed
relative to USNO-B using 60 sources in common between our summed image
and the catalog resulting in an rms positional uncertainty of $0.18''$
in each coordinate.  The position of the afterglow is $\alpha=21^{\rm
  h} 54^{\rm m} 48.63^{\rm s}$, $\delta=+16^{\circ} 53' 27.4''$ (J2000).

We continued to monitor the afterglow in the $r',i',z'$ bands for a
period of about nine days.  A log of all Gemini observations is
provided in Table~\ref{tab:gemini}.  All observations were reduced using
the standard {\tt gmos} package in IRAF.

\begin{figure}
\vspace{-1cm}
\plotone{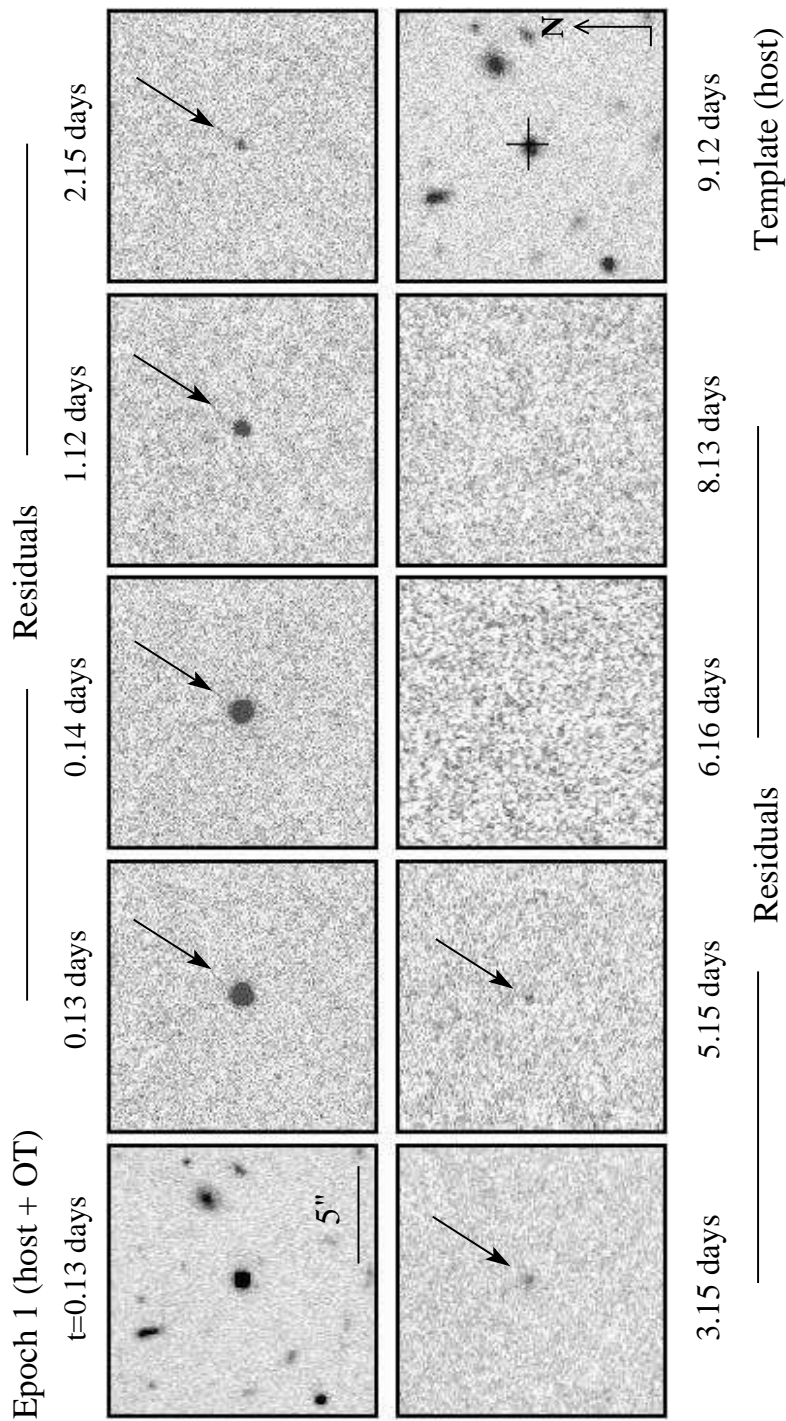}
\vspace{-1.5cm}
\caption{Residual images of the optical afterglow ($r'$-band) of 
\grb\ produced through image subtraction techniques.  The host
galaxy emission dominates the optical light after $t\approx 2$ d.
We assume that the afterglow emission is negligible in the final 
(template) epoch.  The position of the afterglow (cross, template
image) is offset by $0.12\pm 0.04$ arcsec with respect to the host
galaxy center.}
\label{fig:residuals_arrow}
\end{figure}

To study the afterglow variability and remove the contribution from
the bright host galaxy, we used the {\it ISIS} subtraction routine by
\citet{a00} which accounts for temporal variations in the stellar PSF.
Adopting the final epoch observations in each filter as templates we
produced residual images (Figure~\ref{fig:residuals_arrow}).
Photometry was performed on the afterglow emission in each residual
image.  To estimate the error introduced by the subtraction procedure,
we used IRAF/{\tt mkobject} to insert fake stars of comparable
brightness to the afterglow at random positions on each image.  The
images were then subtracted and the false stellar residuals were
photometered in the same manner as the afterglow residual.  We adopt
the standard deviation of the false residual magnitudes as an estimate
for the error introduced by the subtraction procedure.  This error is
negligible for all observations before $t\sim 3$ days and is about
$0.2$ magnitudes for subsequent observations.

For the later epochs in which the afterglow is not detected in our
residual images we estimate our detection threshold by performing
photometry in blank apertures placed at random positions in the
residual images.  We use the standard deviation of the resulting
photometry to estimate the limit ($5\sigma$) to which we could
reliably recover faint sources.

Field calibration was performed for the $r'$-band images using a GMOS
observation of the standard star SA92-250 in photometric conditions on
2005 Dec 30 UT.  To calibrate the $i'$- ($z'$-)band fields we used 23
(26) isolated, unsaturated stars in common with the USNO-B catalog and
converted them to the SDSS photometric system according to the
prescription of \citet{stk+02}.  The photometric uncertainty in this
calibration is 0.20 and 0.40 mag, respectively.  We repeated this
exercise for the $r'$-band to further assess the accuracy of the
USNO-B calibrations and find a zeropoint consistent to $\sim 0.20$ mag
with the standard star calibration.  We estimate our total uncertainty
for the afterglow observations by combining (in quadrature) the errors
associated with the field calibration, subtraction procedure and
photometric measurement (Table~\ref{tab:gemini}).

\begin{figure}
\plotone{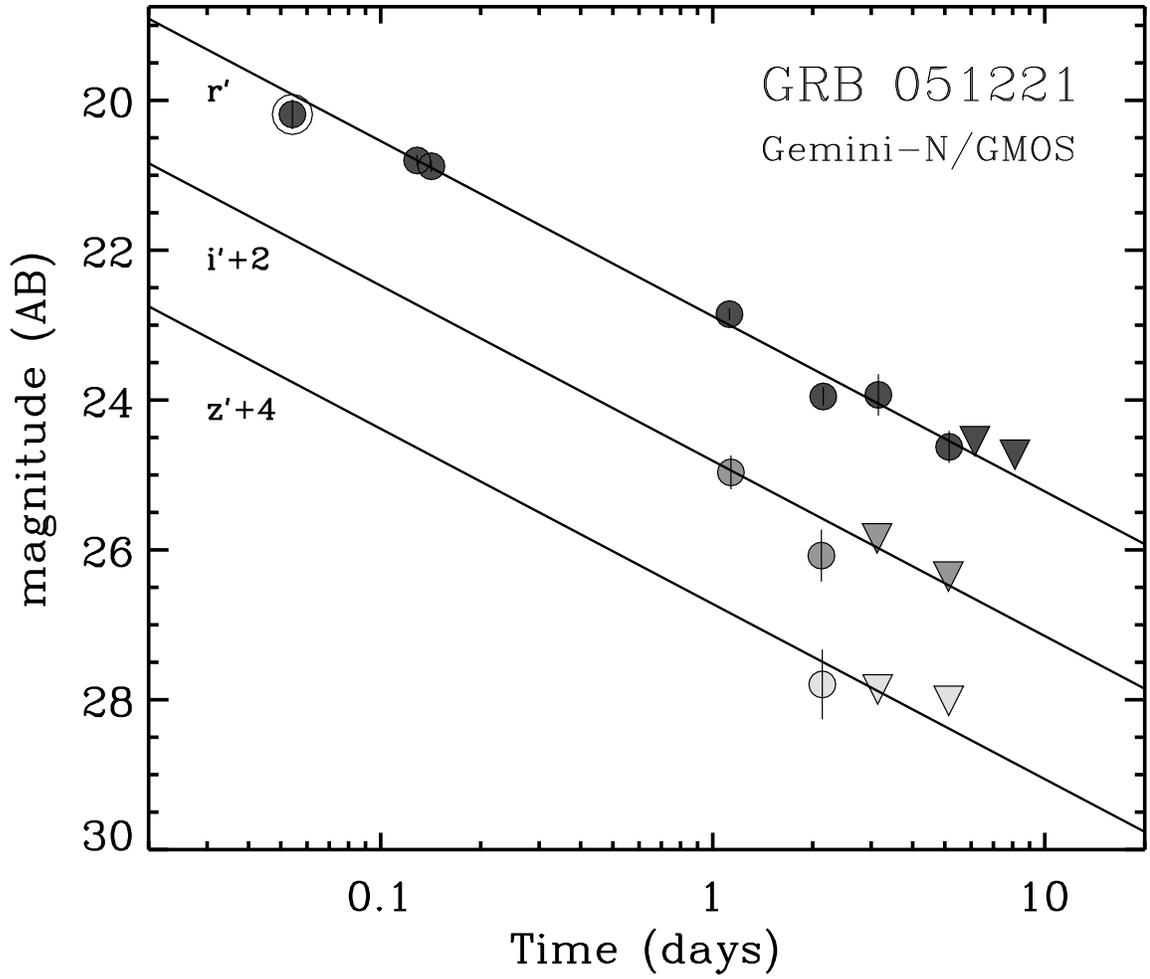}
\caption{Optical ($r'i'z'$) light-curves of the afterglow of \grb\ .
  We have converted the early $R$-band detection by \citet{wvw+05} at
  $t=0.0542$ days to the $r'$-band (encircled dot) using the standard
  zeropoints.  The data have been corrected for Galactic extinction
  of $E(B-V)=0.069$ mag \citep{sfd98}.  The temporal evolution follows a
  decay index of $\alpha_{\rm opt}=-0.92\pm 0.04$. }
\label{fig:opt_lt_curve}
\end{figure}

As shown in Table~\ref{tab:gemini} and Figure~\ref{fig:opt_lt_curve},
the afterglow is clearly detected in our $r'$-band images between
$t\sim 0.1$ and 5.1 days after the burst.  In the first image we find
the afterglow to be $r'=20.99\pm 0.08$ mag (AB system; not including
Galactic extinction of $E(B-V)=0.069$; \citealt{sfd98}) By comparing
the position of the afterglow (as measured in the residual images)
with the center of the host galaxy in the template images, we estimate
an offset of $0.12\pm 0.04$ arcsec (Figure~\ref{fig:residuals_arrow}).

\subsection{Spectroscopic Observations}
\label{sec:spec}

Following the identification of the afterglow in our Gemini images, we
used GMOS to obtain two 1200 s spectra using the R400+G5305 grating
with a central wavelength of 6050 \AA\ and a $1''$ slit.  The data
were obtained on December 22.22 UT, about 27.5 hr after the burst.
The effective wavelength coverage of our spectrum is about $5000-8170$
\AA\ with a resolution of 1.356 \AA/pix ($2\times 2$ binning), and a
resolution element of about 3 pixels.  A second observation consisting
of two 1800 s spectra with the same setting was obtained on December
31.24 UT ($t\approx 10.2$ days).  All observations were reduced using
the standard {\tt gmos} reduction tasks in IRAF before subtracting the
sky, and combining and extracting the spectra using custom packages as
described in \citet{kel03}.  The rms wavelength scatter is about
0.6\AA.

Flux calibration was performed using an archival observation of the
spectrophotometric standard star BD+28$^\circ$\,4211 obtained on 2005
November 17 UT.  The observation was taken with the same instrumental
setup as our observation of \grb\ with the exception of a slight shift
in the central wavelength of $-50$\AA.  This shift was taken into
account in applying the wavelength calibration and the associated
error is likely smaller than that due to slit losses.  Based on our
afterglow photometry (\S\ref{sec:optical}) and a comparison of the two
spectra, we estimate that the afterglow accounts for about $40\%$ of
the total flux in the first spectrum.

\begin{figure}
\plotone{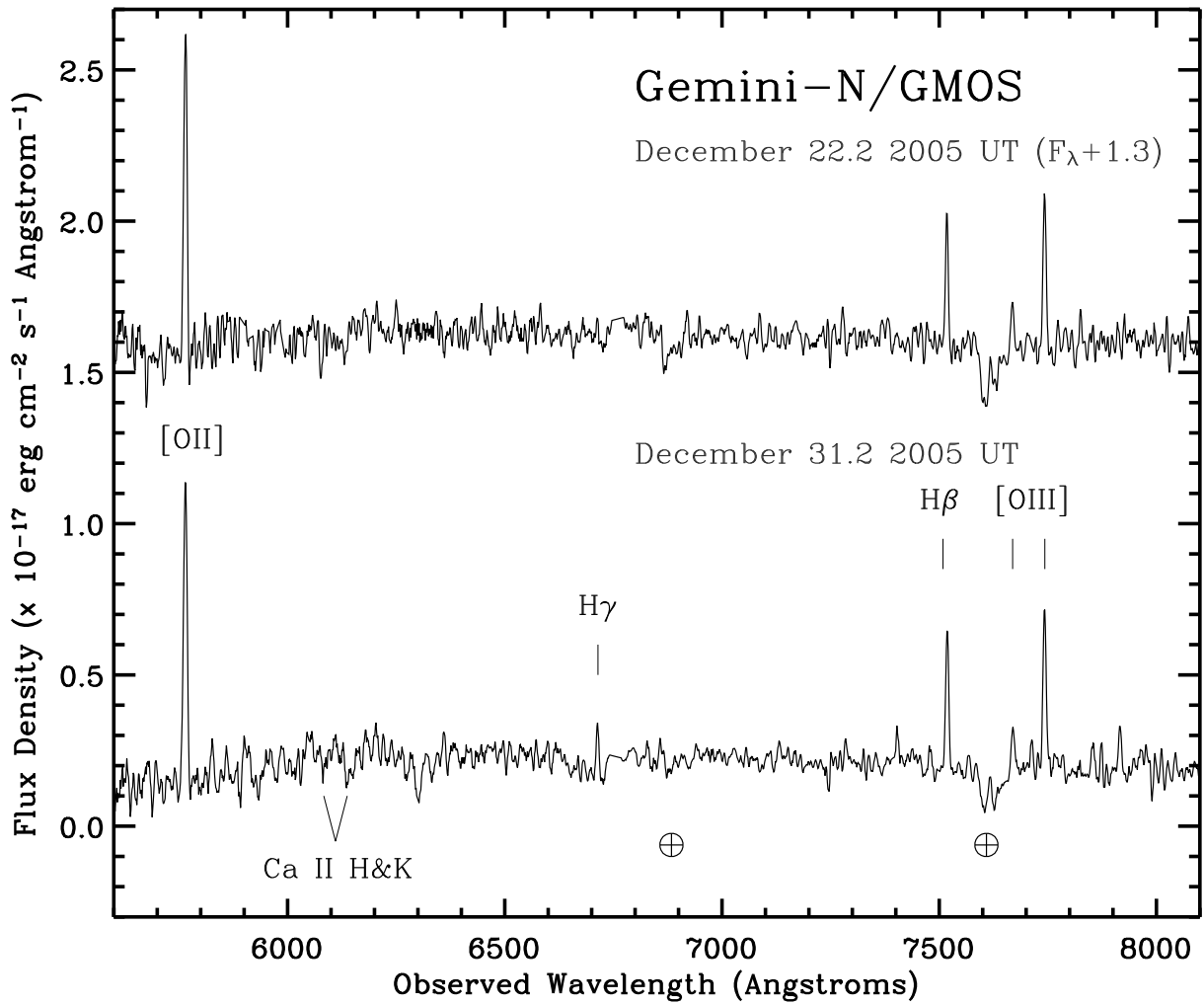}
\caption{Spectrum of the host galaxy of \grb\ taken with Gemini-N/GMOS
  on 2005 December 22.2 and 31.2 ($t\approx 1.1$ and 10.2 days).  We
  detect several bright emission lines which indicate a redshift of
  $z=0.5464\pm 0.0001$ (Table~\ref{tab:lines}).  Absorption features
  attributed to CaII H\&K are also weakly detected.  These features
  indicate that the host is an active star-forming galaxy with a
  significant population of older stars. }
\label{fig:spec_plot}
\end{figure}

As shown in Figure~\ref{fig:spec_plot} and Table~\ref{tab:lines}, we
detect several emission lines in the spectrum from H$\beta$,
[\ion{O}{2}]$\lambda 3727$, and [\ion{O}{3}]$\lambda\lambda 4959,5006$
at a redshift of $z=0.5464\pm 0.0001$.  We detect in addition weak
absorption from CaII H\&K at a redshift $z=0.5458$.  Adopting the
standard cosmological parameters ($H_0=71~\rm km~s^{-1}~Mpc^{-1}$,
$\Omega_M=0.27$, $\Omega_{\Lambda}=0.73$) and a redshift of $z=0.5464$
for the burst, the isotropic gamma-ray energy release is
$E_{\gamma,\rm iso}=4\pi F_{\gamma}d_L^2 (1+z)^{-1}\approx
2.4^{+0.1}_{-1.3}\times 10^{51}$ erg (20 keV -- 2 MeV) where $d_L$ is
the luminosity distance.  Compared with other short-hard bursts
\citep{ffp+05}, the prompt energy release of \grb\ is a factor of
$\sim 6 - 35$ times higher.

\subsection{X-ray Observations}
\label{sec:xray}

\begin{figure}
\plotone{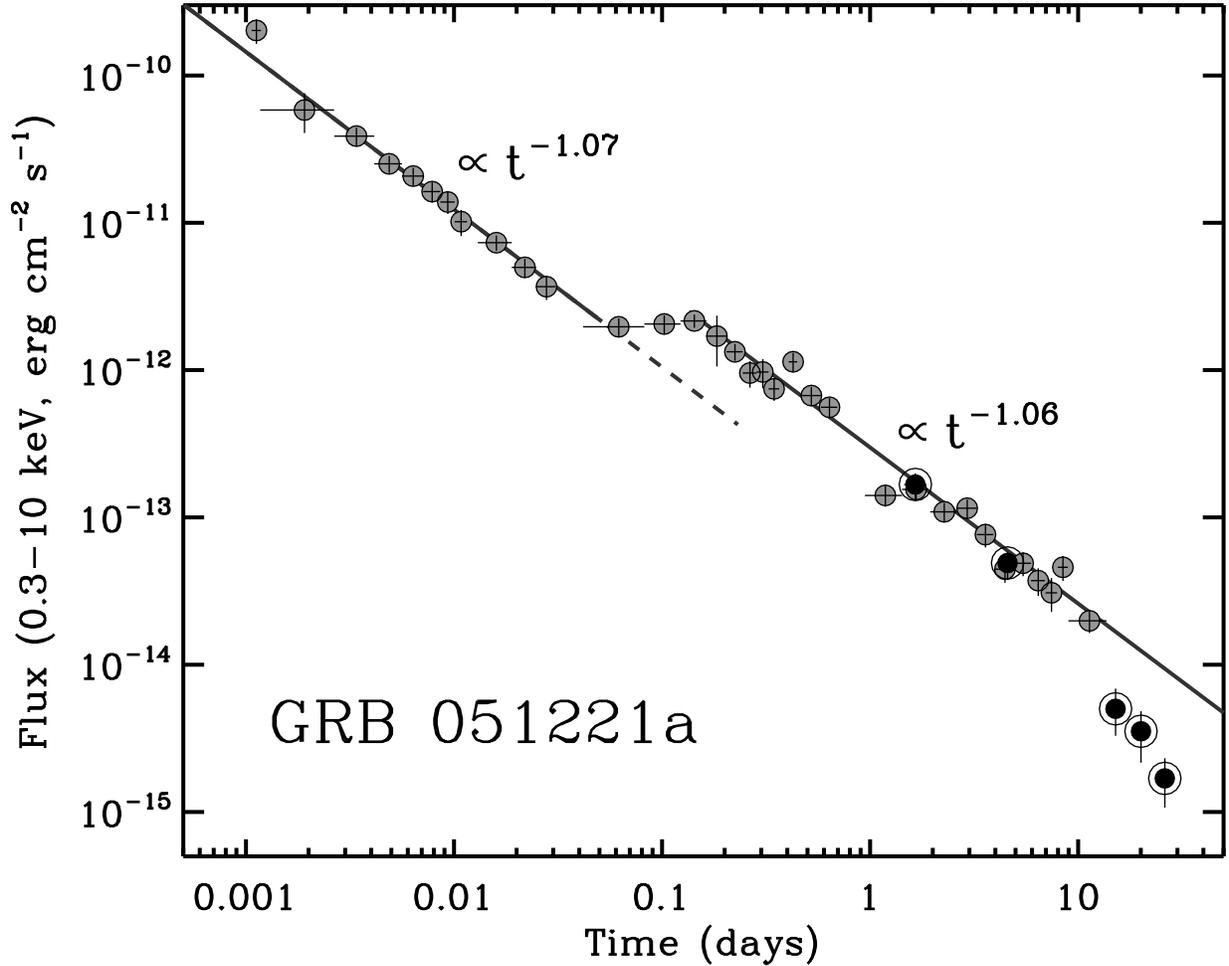}
\caption{Light-curve of the X-ray afterglow of \grb\ as observed with
  \swift/XRT (grey circles) and {\it Chandra}/ACIS-S (encircled black
  dots; \citealt{bgc+06}) from $t\approx 92$ sec to 26 days after the
  burst.  The light-curve is characterized by four phases: an initial
  decay with index $\alpha_X= -1.07\pm 0.03$, a brief plateau phase, a
  subsequent decay with $\alpha_X= -1.06\pm 0.04$ and steepening at
  $t\approx 5$ days.  An extrapolation of the early decay to $t\sim 1$
  day indicates an increase in flux by a factor of $\sim 3.3$
  corresponding to an energy injection of a factor of $\sim 3.4$.  We
  interpret the late-time steepening as a jet break and derive a
  collimation angle of $\theta_j\approx 7$ degrees.}
\label{fig:xrt_lt_curve}
\end{figure}

The X-ray afterglow of GRB\,051221a was monitored with XRT from
$t\approx 92$ s to about 11 days after the burst.  We retrieved the
XRT data from the HEASARC
archive\footnote{http://heasarc.gsfc.nasa.gov/W3Browse/} and performed
the reduction with the {\tt xrtpipeline} script packaged within the
HEAsoft software.  We used the default grade selections and screening
parameters to produce light-curves for the 0.3 -- 10 keV energy range.
Data were re-binned using {\tt lcurve} to produce the final afterglow
light-curve (Figure~\ref{fig:xrt_lt_curve} and Table~\ref{tab:xrt}).
During the first $\sim 17$ hr, the afterglow spectrum is well fit by a
power-law with $\beta_X\approx -1.0\pm 0.2$ ($F_{\nu,X}\propto
\nu^{\beta_X}$), absorbed by a neutral hydrogen column density
$N_H\approx 1.6\pm 0.6\times 10^{21}~\rm cm^{-2}$.  This value is
somewhat larger than that inferred from the Galactic extinction
assuming the standard conversion of
\citet{ps95}.  These parameters are consistent with the analysis by
\citet{bgc+06}.

We use the spectral parameters derived above to calibrate the
light-curve; we find a conversion rate of 1 cps=$5.27\times
10^{-11}~\rm erg~cm^{-2}~s^{-2}$ (0.3--10 keV).  We find that the
afterglow flux at $t=10$ hrs is $F_X\approx 7.5\times 10^{-13}~\rm
erg~cm^{-2}~s^{-1}$ which translates to a luminosity of
$L_{X,10}\approx 5.6\times 10^{44}~\rm erg~s^{-1}$. This value is
one to two orders of magnitude higher than for previous short bursts
\citep{ffp+05} and is in the range typical of the long-duration events
\citep{bkf03}.  Our analysis of the full XRT dataset between $t\approx
92$ sec and 11 days shows that the evolution of the X-ray afterglow
can be characterized by three phases: an initial fading as
$F_{X}\propto t^{\alpha_X}$ with $\alpha\approx -1.07\pm 0.03$, a
brief plateau phase between $t\sim 0.05$ and 0.13 days, and a
subsequent decay as $\alpha_X\approx -1.06\pm 0.04$.  We emphasize
that the decay indices before and after the plateau phase are
identical suggesting a simple evolution interrupted by an injection of
energy (\S\ref{sec:reverse}).

\subsection{Radio Observations}
\label{sec:radio}

We initiated radio observations of GRB\,051221a with the Very Large
Array\footnote{The Very Large Array and Very Long Baseline Array are
  operated by the National Radio Astronomy Observatory, a facility of
  the National Science Foundation operated under cooperative agreement
  by Associated Universities, Inc.} (VLA) on 2005 December 21.99 UT.
The data were obtained at 8.46 GHz covering the range 0.91 -- 23.93
days (see Table~\ref{tab:vla}).  All observations were taken in
standard continuum observing mode with a bandwidth of $2\times 50$
MHz. We used 3C48 (J0137+331) for flux calibration, while phase
referencing was performed against calibrator J2152+175.  The data were
reduced using standard packages within the Astronomical Image
Processing System (AIPS).

We detect a radio source coincident with the optical position with a
flux $F_{\nu}=155\pm 30~\mu$Jy in the first observation.  The source
subsequently faded below our detection threshold with a decay rate of
$\alpha_{\rm rad} \lesssim -1.0$.  

\section{Afterglow Properties}
\label{sec:afterglow}

Using the detailed multi-frequency observations of the \grb\ afterglow
we proceed to constrain the physical properties of the ejecta and the
circumburst density.  We assume that like the afterglows of long GRBs,
the afterglows of short bursts are produced through the dynamical
interaction of the ejecta with the surrounding medium (the forward
shock; FS) with an additional component from shock heating of the
ejecta (the reverse shock; RS).  This assumption is consistent with
the existing afterglow data for short bursts
\citep{bpc+05,bpp+05,ffp+05,p05}.  In this scenario, the total energy
density is partitioned between relativistic electrons, $\epsilon_e$,
and magnetic fields, $\epsilon_B$, while the thermal energy of the
shocked protons accounts for the fraction remaining (see
\citealt{pir99} for a review).  The shocked electrons are accelerated
into a power-law distribution, $N(\gamma)\propto \gamma^{-p}$ above a
minimum Lorentz factor, $\gamma_m$.  The emission resulting from the
forward and reverse shock components is described by a synchrotron
spectrum characterized by three break frequencies --- the
self-absorption frequency, $\nu_a$, the characteristic frequency,
$\nu_m$, and the cooling frequency, $\nu_c$ --- and a flux
normalization, $F_{\nu_m}$ \citep{spn98}.  In modeling the afterglow
spectral and temporal evolution, we adopt the formalism of
\citet{gs02} for a relativistic forward shock expanding into a
constant density circumburst medium and the scalings of \citet{sp99}
for a mildly-relativistic reverse shock.

\subsection{Preliminary Constraints}
\label{sec:prelim}

We proceed to fit the forward shock model to the afterglow data using
only observations at $t\gtrsim 0.1$ day in the X-rays and $t\gtrsim 1$
day in the radio when the afterglow evolution follows a simple
power-law.  As shown in \S\ref{sec:reverse} the data prior to $t\sim
0.1$ day in the X-ray band are influenced by energy injection while
the radio detection at $t\approx 0.91$ days and the subsequent decline
are inconsistent with a simple forward shock model and are instead
explained in the context of reverse shock emission.

To constrain the spectrum of the forward shock, we first investigate
the evolution in the optical and X-ray bands.  From the X-ray data we
measure $\alpha_X=-1.06\pm 0.04$ and $\beta_X=-1.0\pm 0.2$, leading to
$\alpha-3\beta/2=0.44\pm 0.23$.  A comparison to the standard closure
relations (see \citealt{spn98}), $\alpha-3\beta/2=0$ ($\nu_m < \nu <
\nu_c$) and $\alpha-3\beta/2=1/2$ ($\nu > \nu_c$) indicates that
$\nu_X > \nu_c$.  These conclusions are supported by the optical to
X-ray spectral slope, $\beta_{OX}=-0.64\pm 0.05$ (after correcting the
optical data for Galactic extinction), which is flatter than $\beta_X$
as expected if $\nu_{\rm opt} < \nu_c < \nu_X$.  In addition, the
optical decay rate, $\alpha_{\rm opt}=-0.92\pm 0.04$, is somewhat
shallower than $\alpha_X$ as expected if $\nu_c$ is between the two
bands.  We note that this also supports our assumption of a constant
density medium since in a wind environment the expected value of
$\alpha_{\rm opt}$ is about $-1.3$, significantly steeper than the
observed value.  Using all the available optical and X-ray
observations we find that $p=2.15\pm 0.10$ and $\nu_c=(2\pm 1)\times
10^{17}$ Hz at $t=1$ day.  Here and throughout this section we have
accounted for the smooth shape of the spectral breaks and we note that
in this context the values of $F_{\nu_m}$ quoted below are the
asymptotic extrapolations.

Next we compare the optical and radio afterglow data to constrain
$\nu_m$ and the peak spectral flux, $F_{\nu_m}$.  The single power-law
optical decay constrains $\nu_m$ to be lower than the optical band
before the first observation at $t=0.0542$ days, and $F_{\nu_m}$ to be
above the flux of the first observation.  Scaling these constraints to
$t=1$ day ($\nu_m\propto t^{-1.5}$ and $F_{\nu_m}\propto t^0$) we find
$\nu_m\lesssim 1.5\times 10^{12}$ Hz and $F_{\nu_m}\gtrsim 74~\mu$Jy.
Furthermore, we note that the two constraints are linked through the relation
$F_{\nu_m}\propto \nu_m^{-(p-1)/2}$ such that lower values of $\nu_m$
imply increasingly higher values of $F_{\nu_m}$.

Given the link between $\nu_m$ and $F_{\nu_m}$ we obtain a lower bound
on $\nu_m$ (and hence an upper bound on $F_{\nu_m}$) from the limits
on the radio flux at $t\gtrsim 1$ day.  In particular, the $3\sigma$
limits of $\lesssim 80$ $\mu$Jy at $t\sim 2 - 7$ days indicate that
$\nu_m \gtrsim 4.0\times 10^{11}$ Hz and $F_{\nu_m}\lesssim 160~\mu$Jy
at $t=1$ day; otherwise the forward shock flux would be brighter than
the observed radio limits.  Combined with the optically-derived
limits, we therefore find the following allowed ranges: $4.0\times
10^{11} \lesssim \nu_m \lesssim 1.5\times 10^{12}$ Hz and $74 \lesssim
F_{\nu_m} \lesssim 160~\mu$Jy at $t=1$ day.

\subsection{Forward Shock Broad-Band Model}
\label{sec:fit}

Adopting these preliminary constraints on $\nu_m$ and $F_{\nu_m}$ and
using the derived value of $\nu_c$, we apply a broad-band afterglow
model fit to the multi-frequency data in order to determine the
physical parameters of the burst.  The four spectral parameters
($F_{\nu_m}$, $\nu_a$, $\nu_m$ and $\nu_c$) are fully determined by
four physical quantities: the kinetic energy of the ejecta, $E_{K,\rm
iso}$, the energy partition fractions, $\epsilon_e$ and $\epsilon_B$,
and the circumburst density, $n$.  Therefore by constraining the four
spectral parameters through broad-band observations, we are able to
determine a unique solution for the four physical parameters.
Although $\nu_a$ is not directly constrained by the observations, we
are able to define a range of reasonable solutions by requiring that
$\epsilon_e$,$\epsilon_B \le 1/3$, which accounts for an equal (or
larger) contribution from shocked protons ($\epsilon_p \ge
1/3$). This requirement excludes unphysical solutions in which the sum of
the contributions from shocked electrons, protons and magnetic fields
exceed the total energy density.  With this constraint on the energy
partition fractions, and the observed estimates for $F_{\nu_m}$,
$\nu_m$ and $\nu_c$ from \S\ref{sec:prelim}, we find the following
ranges for the physical parameters:
\begin{mathletters}
\begin{eqnarray}
1.1\times 10^{51}\lesssim E_{K,\rm iso}\lesssim 4.4 \times 10^{51}~~\rm erg  
\label{eqn:constr1} \\
2.8\times 10^{-4}\lesssim n\lesssim 1.5\times 10^{-1}~~\rm cm^{-3}
\label{eqn:constr2} \\
9.5 \times 10^{-2}\lesssim\epsilon_e\le 1/3 
\label{eqn:constr3} \\
7.8\times 10^{-3}\lesssim\epsilon_B\le 1/3.  
\label{eqn:constr4}
\end{eqnarray}
\end{mathletters}

\subsection{Energy Injection and Reverse Shock Model}
\label{sec:reverse}

We now turn to the X-ray plateau phase observed between $t\approx 1$
and 3 hrs after the burst.  The flattening of the light-curve suggests
that \grb\ experienced an episode of energy injection on this
timescale.  As shown in Figure~\ref{fig:xrt_lt_curve}, an
extrapolation of the early decay lies a factor of about $3.3$ below
the observed flux after the plateau phase.  Given that $\nu_c$ is
located below the X-ray band on this timescale we have $F_{\nu,X}\propto
E^{(p+2)/4}$.  Thus for $p=2.15$ the flux increase corresponds to an
energy injection of a factor of $\sim 3.4$.  We note that given the
flat X-ray evolution during the energy injection phase the ejecta
refreshing the shock follows the relation $E(>\gamma) \propto \gamma^{-4.5}$
\citep{sm00}.

One implication of this energy injection episode is emission from the
associated reverse shock \citep{sm00}.  The broad-band spectrum of the
reverse shock differs from that of the forward shock spectrum
primarily due to its lower characteristic frequency,
$\nu_m^{RS}\approx \nu_m^{FS}/\gamma^2$, and its higher spectral peak
flux, $F_{\nu_m}^{RS}\approx \gamma F_{\nu_m}^{FS}$.  Here, the bulk
Lorentz factor of the ejecta is given by $\gamma\approx 8.3~(E_{K,\rm
  iso}/10^{51}~{\rm erg})^{1/8} (n/10^{-3}~{\rm cm^{-3}})^{-1/8}
(t/1~\rm day)^{-3/8}$.

Adopting the constraints for $\nu_m^{FS}$ and $F_{\nu_m}^{FS}$ (along
with the associated constraints on $E_{K,\rm iso}$ and $n$) from
\S\ref{sec:afterglow}, we find that at the time of the energy injection
episode ($t\sim 0.1$ days) $18 \lesssim \gamma \lesssim 26$ and thus
$2.0\times 10^{10} \lesssim \nu_m^{RS}\lesssim 1.5\times 10^{11}$ Hz
and $1.3 \lesssim F_{\nu_m}^{RS} \lesssim 4.1$ mJy.  Using the
temporal scalings of \citet{sp99}, $\nu_m^{RS}\propto t^{-73/48}$ and
$F_{\nu_m}^{RS}\propto t^{-47/48}$ we find that $\nu_m^{RS}$ is just
below 8.46 GHz during our first radio observation. We therefore
estimate $47 \lesssim F_{\nu,\rm 8.46~GHz}^{RS} \lesssim 295~\mu$Jy at
$t\approx 0.91$ days which is fully consistent with our measured flux of
$F_{\nu,\rm 8.46}=155\pm 30$.

The radio RS component is then predicted to decay rapidly as
$F_{\nu}\propto t^{-1.85}$ implying an estimated flux at $t= 1.94$
days (our second radio epoch) of $12 \lesssim F_{\nu,\rm
  8,46~GHz}^{RS} \lesssim 73~\mu$Jy, again consistent with our
measured limit of $F_{\nu,\rm 8.46}\lesssim 72~\mu$Jy.  We therefore
conclude that the X-ray plateau phase at $t\sim 0.1$ day and the early
radio decay are the result of an energy injection episode.  We note
that this interpretation predicts a 4.86 GHz flux density at $t\approx
0.6$ days that is a factor of $\sim 2$ above the measured limit of
\citet{v05}.  However, this can be explained as the result of
synchrotron self-absorption since for the reverse shock,
$\nu_a^{RS}\approx \gamma^{8/5}\nu_a^{FS}$, can be as high as 5 GHz
at $t=0.6$ days given our allowed range of $\nu_a^{FS}$.

\subsection{A Combined FS+RS Model}
\label{sec:sum}

\begin{figure}
\plotone{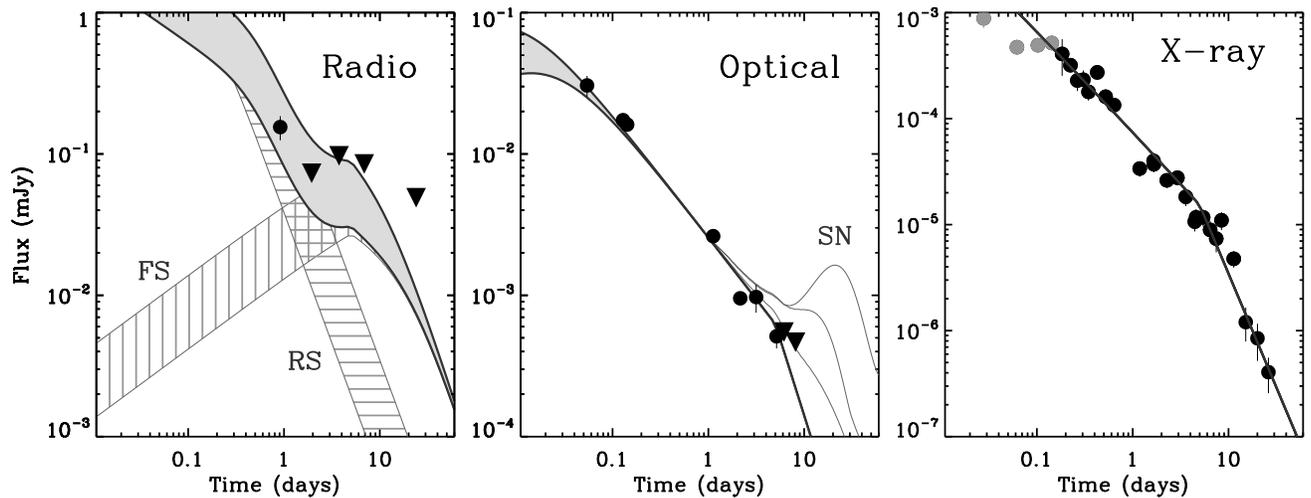}
\caption{Broad-band afterglow model fits for the radio (8.5 GHz), optical
  ($r'$-band) and X-ray data of \grb\ based on the inferred
  constraints discussed in \S\ref{sec:fit}.  Detections are shown as
  filled circles and upper limits are plotted as inverted triangles.
  The $r'$-band data have been corrected for Galactic extinction.
  Black lines designate the forward shock contribution while the
  shaded region in the radio includes the allowed range of
  $\nu_m^{FS}$ and $F_{\nu_m}^{FS}$ (see also optical band) and the
  reverse shock emission associated with the energy injection episode
  in the X-rays at $t\sim 0.1$ days.  Overplotted in the optical panel
  are the predicted light-curves including the contribution from Type
  Ibc SNe 1998bw, 1994I, and 2002ap (thin grey lines; top-to-bottom,
  respectively).  The late-time optical limits clearly rule out the
  presence of a bright SN\,1998bw-like event and constrain the peak
  optical magnitude to $M_V\gtrsim -17.2$ mag (rest-frame),
  consistent with a faint event similar to SN\,2002ap.  At jet break
  is observed at $t\approx 5$ days, causing an achromatic steepening
  of the afterglow light-curves.}
\label{fig:model}
\end{figure}

In Figure~\ref{fig:model} we show the combined FS+RS light-curves and
we find that given the contribution of the reverse shock emission in
the radio band we can limit the range of allowed physical parameters
beyond the constraints provided in
Equations~\ref{eqn:constr1} to \ref{eqn:constr4}.  Using the observed
upper limits, we therefore favor the fainter end of the allowed FS
model corresponding to the parameters $\nu_m^{FS}\approx 1.5\times
10^{12}$ Hz and $F_{\nu_m}^{FS}\approx 74~\mu$Jy at $t=1$ day.  This
provides the following constraints on the physical parameters:
$E_{K,\rm iso}\approx (1.1-1.6)\times 10^{51}$ erg, $n\approx (0.5-2.4)\times
10^{-3}$ cm$^{-3}$, $\epsilon_e\approx (0.24-0.33)$ and
$\epsilon_B\approx (0.12 - 0.33)$.

These improved constraints lead to the following implications. First,
the efficiency in converting the energy in the ejecta into
$\gamma$-rays is $\eta_\gamma\approx 0.6-0.7$ indicating a roughly
equal partition in prompt and afterglow energy, similar to what is
found for long GRBs \citep{pk02}.  Second, we find that the shock
energy is in rough equipartition between relativistic electrons and
magnetic fields.  This is consistent with the values inferred
for long-duration GRBs \citep{yhs+03,pk02}.

\subsection{A Jet Break}
\label{sec:jet}

Using late-time {\it Chandra} ACIS-S observations extending to 26
days, \citet{bgc+06} find strong evidence for a steepening in the
X-ray afterglow light-curve, which they attribute to a jet break.  By
supplementing our optical,XRT and radio data with the late-time {\it
Chandra} observations, we are able to constrain the opening angle of
the jet and thus derive the beaming-corrected energy of the burst.  As
shown in Figure~\ref{fig:xrt_lt_curve}, the first two {\it Chandra}
epochs at $t\approx 1.7$ and 4.6 days are consistent with the {\it
Swift}/XRT measurements and the inferred decay of $\alpha_X\approx
-1.06$.  However, the last three {\it Chandra} points at $t\gtrsim 15$
days show a steeper decay, $\alpha_X=-2.0\pm 0.4$, which is consistent
with the expected decay rate in the case of a jet break,
$\alpha_X=-p\approx -2.15$ \citep{sph99}.  By fitting the radio,
optical and X-ray data together, we modify our spherical FS+RS model
to include a jet break and find a best fit for a smooth break at
$t_j\approx 5$ days, consistent with the findings of \citet{bgc+06}.
As shown in Figure~\ref{fig:model}, this model provides an excellent
fit to the entire broad-band data set.

Together with the physical parameters derived in \S\ref{sec:sum}, the
observed jet break constrains the collimation of the ejecta to
$\theta_j\approx 6.6 (E_K/10^{51}~{\rm erg})^{-1/8} (n/10^{-3}~{\rm
cm^{-3}})^{1/8}\approx (5.7 - 7.3)$ degrees. This indicates that the
true energy release is $E_\gamma\approx (1.2 - 1.9)\times 10^{49}$ erg
and $E_K\approx (7.8 - 8.9)\times 10^{48}$ erg, or a total
relativistic energy yield of $(E_{\gamma}+E_K)\approx (2 - 3)\times
10^{49}$ erg.  We summarize the final parameters for \grb\ in
Table~\ref{tab:params} and note that they are generally consistent
with those reported by \citet{bgc+06}.  Finally, we find that the
ejecta will become non-relativistic on a timescale $t_{NR}\approx 2.0
(E_K/10^{51})^{1/3} (n/10^{-3}~\rm cm^{-3})^{-1/3}\approx (0.3-0.5)$
yrs.

\subsection{Limits on an Associated Supernova}
\label{sec:sn}

The optical afterglow measurements at late-time can also be used to
constrain the contribution from an associated supernova. To assess
this contribution we adopt the optical data for the local Type Ibc SNe
1994I \citep{rvh+96}, 1998bw \citep{gvv+98,ms99} and 2002ap
\citep{fps+03} as templates.  These three SNe were selected based on
their well-sampled optical light-curves and peak luminosities which
represent the overall spread in the observed properties of Type Ibc
supernovae.  To produce synthesized light-curves for each of these
template SNe, we compiled optical {\it UBVRI} observations from the
literature and smoothed the extinction-corrected (foreground plus host
galaxy) light-curves.  We then produced k-corrected light-curves by
redshifting the interpolated photometric spectrum and stretching the
arrival time of the photons by a factor of $(1+z)$.  A full discussion
of the template datasets and the method by which we produce the
synthetic SN light-curves appears in \citet{skf+05}.

Shown in Figure~\ref{fig:model} are the synthesized $r'$-band
light-curves for these three supernovae at the redshift of \grb\
(roughly equivalent to the rest-frame $B$-band SN emission).  These
light-curves represent the summed contribution from the afterglow and
SN models.  Clearly, our late-time optical measurements rule out the
presence of the brighter two SNe (1998bw and 1994I) while our limits
are consistent with a faint SN\,2002ap-like event with a peak optical
magnitude of $M_V\gtrsim -17.2$ (rest-frame) corresponding to a limit
on the mass of synthesized Nickel of $\lesssim 0.07~M_{\odot}$
\citep{mdm+02}.  We note, however, that the bright afterglow flux at
$t\lesssim 1$ day can easily hide the emission from a mildly
relativistic ``macronova'' for a wide range of ejecta velocities and
energies \citep{k05}, and we therefore cannot assess the contribution
from such a component. Finally, we note that no SN absorption features
were observed in the spectrum taken at $t\approx 10.2$ days at which
time a typical Type Ibc SN would reach maximum light.

\section{Host Galaxy Properties}
\label{sec:host}

We now turn to the properties of the GRB host galaxy.  We measure the
brightness of the host galaxy using our final epoch template images
(Table~\ref{tab:gemini}) and find $r'=21.99\pm 0.09$ mag,
$i'=21.99\pm 0.17$ mag, and $z'= 21.97\pm 0.40$ mag; all magnitudes
are in the AB system, have been corrected for Galactic extinction
$E(B-V)=0.069$ mag, and the uncertainties are dominated by the photometric
calibration.  The spectral slope based on these magnitudes is nearly
flat in $F_\nu$.  At the redshift of the host galaxy the rest-frame
$V$-band is traced by the combination of the $i'$ and $z'$ bands,
leading to an absolute magnitude, $M_V\approx -20.0\pm 0.3$ mag, while
the rest-frame $B$-band is traced by the $r'$-band indicating
$M_B\approx -19.9\pm 0.1$ mag, or a luminosity $L_B\approx 0.3L^{*}$.
This luminosity is similar to those of the host galaxies of many long
GRBs, as well as the host galaxy of the short GRB\,050709 with
$L\approx 0.1L^{*}$ \citep{ffp+05}.

We further use {\tt galfit} \citep{phi+02} to examine the surface
brightness profile of the host galaxy.  We use nearby stars to
construct a point-spread function image for the deconvolution process
and find that the best-fit scale length is about $420\pm 50$ mas, or
about $2.6\pm 0.3$ kpc, and the Sersic index is about 0.6, close to
the value of 1 for an exponential disk, but significantly different
from a value of 4 appropriate for elliptical galaxies. The
morphological parameters are again similar to those found for the
hosts of long-duration GRBs \citep{wbp05}.

At $z=0.5464$, the measured offset of the afterglow
(\S\ref{sec:optical}) relative to the center of the host galaxy
corresponds to $760\pm 30$ pc.  Normalized by the scale length of the
galaxy this corresponds to $r/r_e=0.29\pm 0.04$.  This offset is
somewhat smaller than those measured for previous short GRBs
\citep{ffp+05} and continues the trend of smaller offsets
compared to predictions from population synthesis models. For example
\citet{fwh99} predict that $<1\%$ of short GRBs will have offsets less
than 2 kpc.  We note, however, that smaller offsets are possible for
merger events within star-forming galaxies, which may have small
initial binary separations \citep{bbk02}.

As shown in Figure~\ref{fig:spec_plot}, the host exhibits both
emission lines and \ion{Ca}{2} H\&K absorption features.  Using our
late-time spectrum in which the afterglow contribution is negligible,
we estimate the star formation rate in the host galaxy from the
observed fluxes of the various emission lines.  From the flux of the
[\ion{O}{2}]$\lambda 3727$ line, $F_{\rm [OII]}=1.3\times 10^{-16}$
erg cm$^{-2}$ s$^{-1}$ (Table~\ref{tab:lines}), and the
conversion of \citet{ken98}, ${\rm SFR}=(1.4\pm 0.4)\times
10^{-41}L_{\rm [OII]}$, we find a star formation rate of about $2.0\pm
0.5$ M$_\odot$ yr$^{-1}$.  From the H$\beta$ line flux, $F_{\rm
  H\beta}\approx 4.5\times 10^{-17}$ erg cm$^{-2}$ s$^{-1}$, and assuming the case-B recombination ratio of $F_{\rm
  H\alpha}/F_{\rm H\beta}=2.85$ we infer a star formation rate of
about $1.2$ M$_\odot$ yr$^{-1}$.  Thus, we conclude that the star
formation rate (not corrected for host extinction) is about $1.6\pm
0.4$ M$_\odot$ yr$^{-1}$.  We note that the measured ratio of
H$\gamma$/H$\beta$ $\sim 0.3$, compared to the theoretical value of
about $0.47$ \citep{ost89}, may point to an extinction correction of
the star formation rate of about a factor of two.  The inferred star
formation rate is not atypical for the hosts of long GRBs, but is
about a factor of six larger than in the star-forming host of the
short GRB\,050709 \citep{ffp+05}.  Moreover, it exceeds by at least a
factor of $20-30$ the limits on the star formation rates in the
elliptical hosts of GRBs 050509b and 050724
\citep{bpc+05,bpp+05,pbc+05}.

The combination of the inferred star formation rate and host
luminosity provides a measure of the specific star formation rate.  We
find a value of $4$ M$_\odot$ yr$^{-1}$ $L$*/$L$, which is about a
factor of 2.5 times lower than the mean specific star formation rate
for the hosts of long-duration GRBs \citep{chg04}. This indicates that
the star formation activity in the host of \grb\ is less intense than
in a typical long GRB host galaxy.

The detection of \ion{Ca}{2} H\&K absorption lines indicates that
while the host is under-going current star formation, at least some
fraction of the light arises from an old population of stars.  The
relative strengths of the two \ion{Ca}{2} lines provide an indication
of the age of the stellar population.  Following the approach of
\citet{ros85} we find that the ratio of \ion{Ca}{2}\,H+H$\epsilon$ to
\ion{Ca}{2}\,K is about unity.  This value is typical for stars with
spectral type later than F; for earlier type stars \ion{Ca}{2}
reversal takes place.  An accurate decomposition of the fractions of
old and new stars is difficult to achieve given that these features
are only weakly detected.  Still, a contribution from a stellar
population with an age of $\sim 1$ Gyr is suggested.

Finally, we use the relative strengths of the oxygen and hydrogen
emission lines to infer the ionization state and oxygen abundance.
The relevant indicators are $R_{23}\equiv {\rm log}\,(F_{\rm
  [OII]}+F_{\rm [OIII]}/F_{\rm H\beta})\approx 0.62$ and $O_{32}\equiv
{\rm log}\,(F_{\rm [OIII]}/ F_{\rm [OII]})\approx -0.24$.  Using the
calibration of \citet{mcg91} we find that the for the upper branch the
metallicity is $12+{\rm log}\,({\rm O/H})\approx 8.7$ while for the
lower branch it is about 8.2; the two branches are due to the
double-valued nature of $R_{23}$ in terms of metallicity.  Thus, the
host metallicity is $0.3-1$ $Z_{\odot}$.  This is somewhat higher
than in the host galaxies of long GRBs, some of which have
metallicities that are $\sim 1/10$ solar (e.g., \citealt{pbc+04}).
This supports the conclusion that the stellar population in the host
galaxy of \grb\ is more evolved than that in the host galaxies of long
GRBs.

To summarize, we find that the host galaxy of \grb\ is undergoing
active star formation with a rate and intensity at the low end of the
distribution for the hosts of long GRBs, but larger than all previous
hosts of short GRBs.  The active star formation with an [\ion{O}{2}]
equivalent width of about 36\AA, signs of an old stellar population
from the \ion{Ca}{2} H\&K absorption, and a non-detected
Balmer/4000\AA\ break indicate a classification intermediate between
e(b) and e(c) in the classification scheme of \citet{dsp+99}.  Along
with a nearly solar metallicity, the host galaxy of \grb\ appears to
have a relatively evolved population of stars.  This continues to
support the trend that the progenitors of short bursts arise from old
stellar populations, but given the higher star-formation rate compare
to previous short burst hosts leaves open the possibility that the
progenitor life-times span a wide range.

\section{Discussion and Conclusions}
\label{sec:disc}

We present broad-band optical, radio, and X-ray data for the afterglow
of the short-hard burst \grb.  Due to our rapid identification of the
afterglow we were able to obtain a spectrum at $t\approx 1.1$ day when
the afterglow was in principle sufficiently bright to detect
absorption features from the interstellar medium of the host galaxy
leading to a direct redshift measurement.  Unfortunately, at the
redshift of the host galaxy, $z=0.5464$, no strong metal UV lines are
redshifted into the waveband of our spectrum.  Still, rapid
identification and spectroscopy of future short bursts may potentially
lead to direct redshift measurements.

At $z=0.5464$, \grb\ currently represents the most distant short GRB
for which we have a secure redshift.  More importantly, we emphasize
that \grb\ is only the third short burst for which a detailed
afterglow study has be performed and the physical parameters have been
constrained.  The isotropic-equivalent $\gamma$-ray and kinetic energy
releases are $2.4\times 10^{51}$ and $1.4\times 10^{51}$ erg
respectively.  These values are a factor of 10 and $10^2$ times larger
than the values for the short GRBs 050724 and 050709, respectively.

\begin{figure}
\plotone{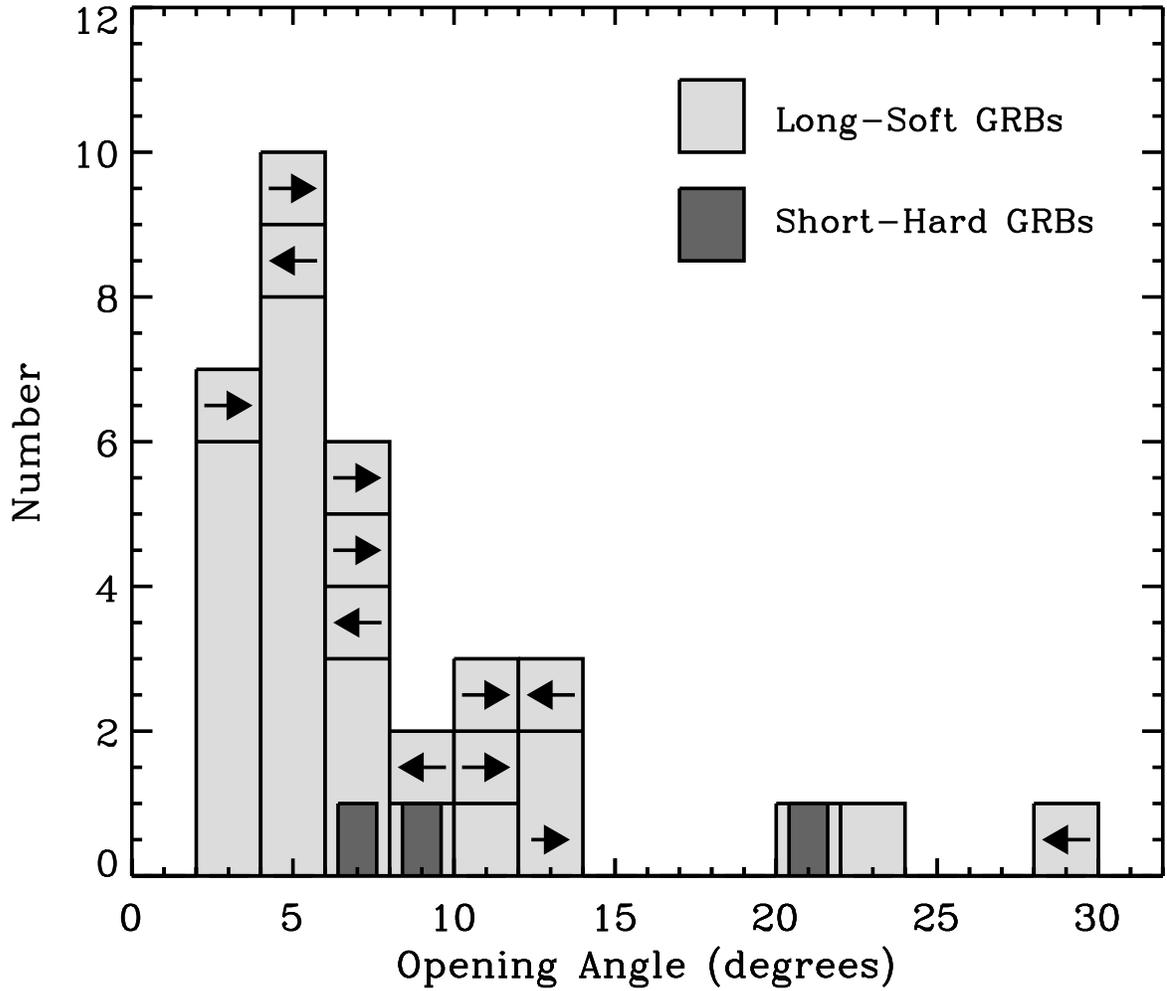}
\caption{The collimation angles for long (\citealt{bkf03,bfk03,ggl04}
  and references therein) and short (GRB\,050709: \citealt{ffp+05};
  GRB\,050724: \citealt{bpc+05}; GRB\,051221a: this paper) GRBs are
  compiled from the literature. All three short bursts have opening
  angles larger than $5$ degrees which is the median value for
  long bursts.  }
\label{fig:angles_hist}
\end{figure}

Adopting the results of our standard synchrotron model and using
$t_j\approx 5$, we constrain the collimation of the ejecta to
$\theta_j\approx 7^{\circ}$, comparable to the opening angles of GRBs
050709 and 050724 (\citealt{ffp+05,bpc+05,p05}; but see \citealt{gbp+06}). A
comparison to the compilation of jet opening angles of long GRBs,
shown in Figure~\ref{fig:angles_hist} demonstrates that while overlap
exists between the two distributions, the opening angles measured for
the three short GRBs are about a factor of two larger than the median
value of $5^{\circ}$ for long GRBs.  This may not be surprising given
that in the case of long GRBs the passage of the jet through the
stellar envelope is thought to be responsible for its collimation
\citep{zwm03}. On the other hand, in models of compact object mergers
it has been argued that the outflow may be collimated by a neutrino
driven wind \citep{rrd03} or the accretion disk itself \citep{ajm05}
both of which lead to relatively wide opening angles, $\gtrsim
10^{\circ}$.  Most likely the same holds true for scenarios such as
accretion induced collapse of a neutron star or a white dwarf
\citep{yb98,fbh+99}.

\begin{figure}
\plotone{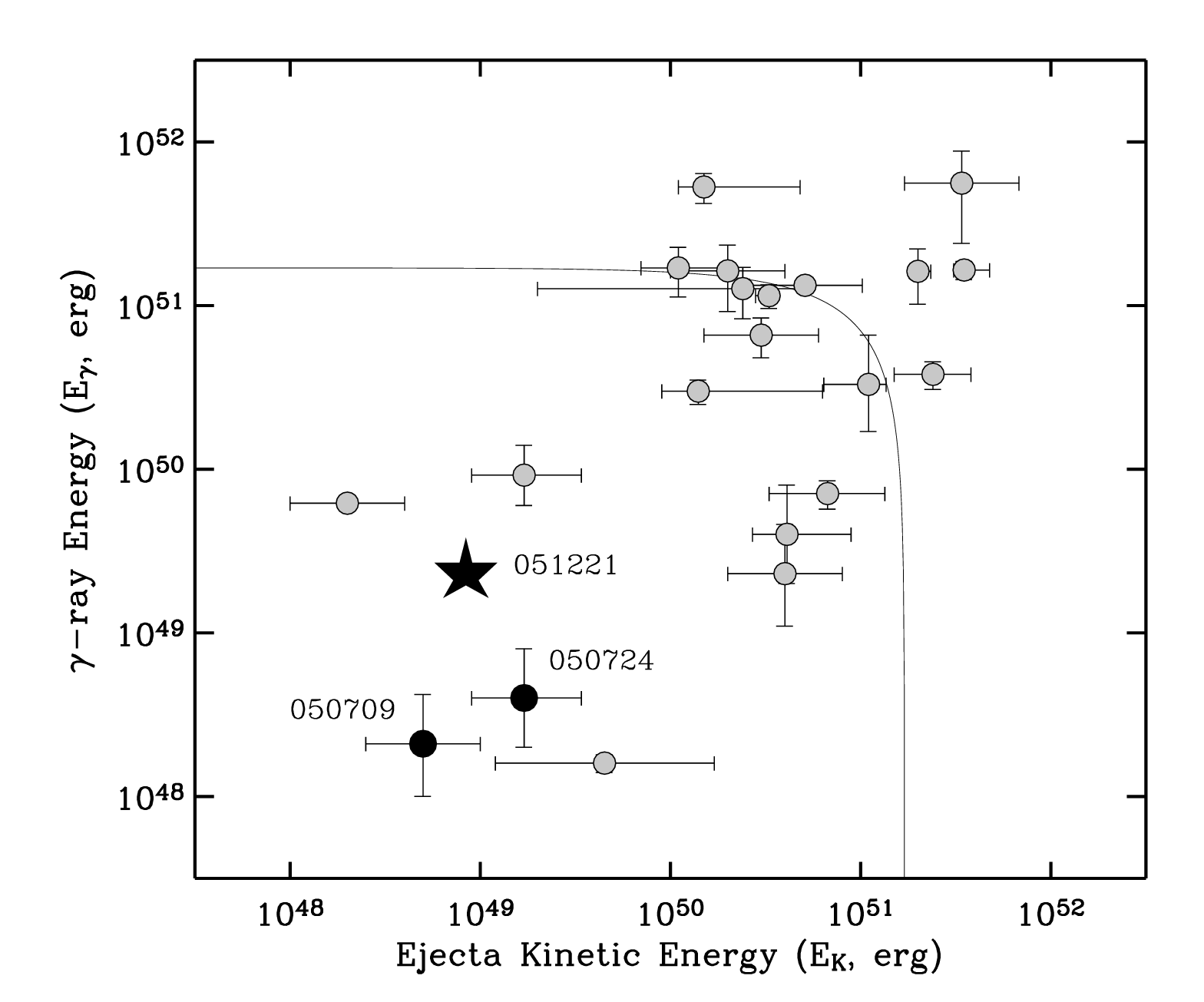}
\caption{The energy release in prompt gamma-ray emission is plotted
  against the total kinetic energy of the ejecta for long GRBs (grey
  circles; \citealt{bfk03,ggl04,sls04,snb+05,skn+06} and references
  therein) and short bursts (black circles; GRB\,050709:
  \citealt{ffp+05}; GRB\,050724: \citealt{bpc+05}) as inferred from
  broad-band afterglow modeling. Both axes are corrected for
  collimation of the ejecta.  The total energy release (prompt plus
  afterglow) of long bursts cluster near $1.7\times 10^{51}$ erg
  (black solid arc). GRB\,051221a (star) has a total energy comparable
  to short GRBs 050724 and 050709.}
\label{fig:egamma_eag}
\end{figure}

With the inferred opening angle, the beaming-corrected energy of \grb\
are similar to those of previous short bursts and comparable to the
low end of the distribution for long duration events (e.g.,
GRB\,031203; \citealt{sls04,skb+04}).  This is shown in
Figure~\ref{fig:egamma_eag} where we compile the beaming-corrected
gamma-ray energy release and ejecta energy values for both long and
short duration bursts.  It has been argued that the majority of the
long GRBs have a total (prompt plus afterglow) true energy release
within $0.5$ dex of $1.7\times 10^{51}$ erg \citep{bkp+03}, with $3$
events having a total energy of $\lesssim 10^{50}$ erg
\citep{skb+04,skn+06}.  The similar $\gamma$-ray efficiencies of long
and short bursts indicate that the energy dissipation process of the
prompt emission is most likely the same.

The energy release inferred for \grb\ can also be used to constrain
the energy extraction mechanism particularly in the context of merger
models.  Two primary mechanisms for energy extraction from the central
engine have been discussed in the literature, namely
$\nu\overline{\nu}$ annihilation or magnetohydrodynamic (MHD)
processes, for example the Blandford-Znajek process (e.g.,
\citealt{bz77,lrp05,rrd03}).  Numerical simulations indicate that
$\nu\overline{\nu}$ annihilation generally produces a less energetic
outflow than MHD processes \citep{lrg05}.  \citet{rr02} show that even
under the most favorable conditions, neutrino annihilation is unlikely
to power a burst with a beaming-corrected energy in excess of few
$\times 10^{48}$ erg, while magnetic mechanisms can produce
luminosities $\gtrsim 10^{52}~\rm erg~s^{-1}$ \citep{rrd03}.  Given
that the beaming-corrected total energy for GRB\,051221a is a factor
of $\sim 10$ larger than the predicted yield for the
$\nu\overline{\nu}$ annihilation model, this may indicate that MHD
processes are responsible.

\begin{figure}
\plotone{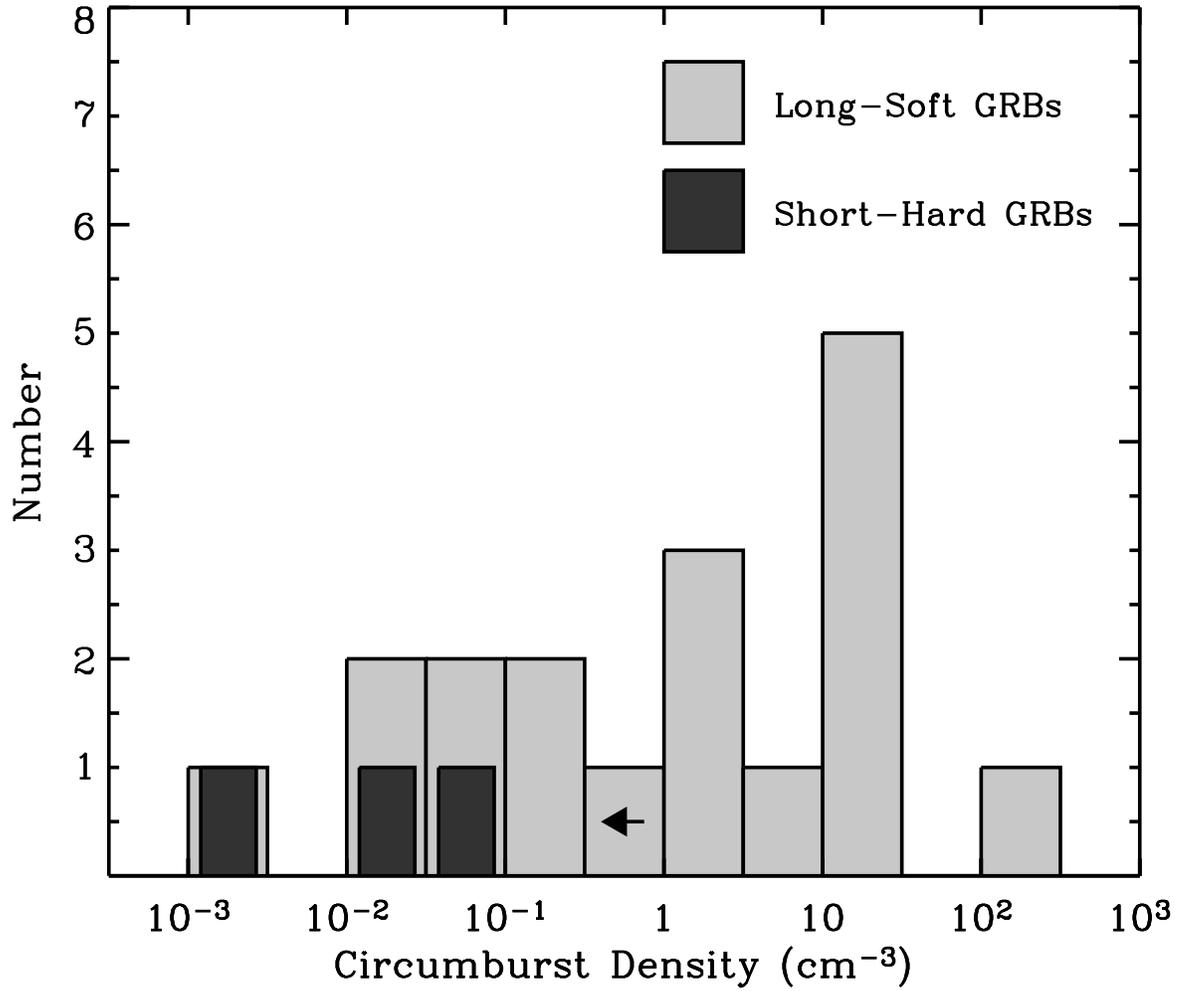}
\caption{The circumburst density values for long (\citealt{snb+05} and
  references therein) and short (GRB\,050709: \citealt{ffp+05};
GRB\,050724: \citealt{bpc+05}; GRB\,051221a: this paper) are shown as a
histogram.  While long bursts show a wide range of
circumburst densities, the short events cluster toward lower
values.  This can be understood in terms of their different
progenitors and environments.}
\label{fig:ed_hist}
\end{figure}

We next compare the circumburst densities of long and short GRBs.  In
Figure~\ref{fig:ed_hist} we show that the circumburst densities for
short bursts cluster near $n\sim 10^{-2}~\rm cm^{-3}$.  On the other
hand, the densities inferred for long GRBs range from about $10^{-2}$
to $10~\rm cm^{-3}$ with a median value of about $2~\rm cm^{-3}$. The
fact that the circumburst densities of short GRBs cluster at the low end 
of the distribution for long GRBs supports the notion that short bursts do
not explode in rich stellar wind environments but are instead consistent with
interstellar densities.

Finally, we show that the host galaxy of GRB\,051221a is actively
forming stars at a rate of about 1.6 M$_\odot$ yr$^{-1}$, but at the
same time exhibits evidence for an appreciable population of old stars
($\sim 1$ Gyr).  The inferred star-formation rate is at least an order
of magnitude larger than the values inferred for previous short GRB hosts
and yet is at the low end of the distribution for long GRB hosts.
Similarly, the inferred metallicity for the \grb\ host galaxy appears
to be higher than those of long GRB hosts indicating that it is more
evolved.  It is conceivable that the spread in short GRB host
properties reflects a dispersion in the progenitor lifetimes such that
the progenitor of \grb\ may be younger than those of GRBs 050509b and
050724 which occurred in elliptical galaxies (see also
\citealt{pbc+05,gno+05}).  The potential spread in progenitor
lifetimes may also be supported by the smaller offset of
\grb\ compared to previous short bursts, but we caution that offsets
are affected not only by progenitor lifetimes but also by the potential
dispersion in kick velocities, host masses, and projection effects.

While the basic properties of short GRBs (redshifts, hosts, isotropic
$\gamma$-ray energies) are now available for several events, it is
clear that detailed afterglow observations are required for a complete
understanding of the burst properties.  With the addition of \grb\ to
the existing sample of only two well-studied events, we have embarked
on a quantitative study of short bursts that over the next few years
will shed light on the diversity of progenitors and energy extraction
mechanisms.

\begin{acknowledgements}
The authors thank S. Rosswog for helpful discussions.  As always, the
authors thank Jochen Greiner for maintaining his GRB page.  A.M.S. and
S.B.C. are supported by the NASA Graduate Student Research Program.
E.B. is supported by NASA through Hubble Fellowship grant
HST-HF-01171.01 awarded by the STScI, which is operated by the
Association of Universities for Research in Astronomy, Inc., for NASA,
under contract NAS 5-26555.  A.G. acknowledges support by NASA through
Hubble Fellowship grant \#HST-HF-01158.01-A awarded by STScI.  K.R. is
supported by the Gemini Observatory, which provided observations
presented in this paper, and which is operated by the Association of
Universities for Research in Astronomy, Inc., under a cooperative
agreement with the NSF on behalf of the Gemini partnership: the
National Science Foundation (United States), the Particle Physics and
Astronomy Research Council (United Kingdom), the National Research
Council (Canada), CONICYT (Chile), the Australian Research Council
(Australia), CNPq (Brazil) and CONICET (Argentina).  GRB research at
Caltech is supported through NASA.
\end{acknowledgements}

\bibliographystyle{apj1b}

\clearpage

\begin{deluxetable}{lrcr}
\tablecaption{Gemini-N/GMOS Observations of GRB\,051221a}
\tablewidth{0pt} \tablehead{
\colhead{Date Obs} & \colhead{$\Delta t$} & \colhead{Filter} & \colhead{magnitude\tablenotemark{a}} \\
\colhead{(UT)} & \colhead{(days)} & \colhead{} & \colhead{(AB)}
}
\startdata
2005 Dec 21.2060 & 0.1287 & $r'$ &  $20.99\pm 0.08$ \\ 
2005 Dec 21.2194 & 0.1421 & $r'$ &  $21.07\pm 0.08$ \\
2005 Dec 22.2000 & 1.1227 & $r'$ &  $23.04\pm 0.08$ \\
2005 Dec 22.2114 & 1.1341 & $i'$ &  $23.11\pm 0.23$ \\
2005 Dec 23.2014 & 2.1241 & $i'$ &  $24.22\pm 0.35$ \\
2005 Dec 23.2150 & 2.1377 & $z'$ &  $23.92\pm 0.40$ \\
2005 Dec 23.2280 & 2.1515 & $r'$ &  $24.14\pm0.12$ \\
2005 Dec 24.2020 & 3.1247 & $i'$ &  $< 23.99$  \\
2005 Dec 24.2158 & 3.1385 & $z'$ &  $<23.98$  \\
2005 Dec 24.2294 & 3.1530 & $r'$ &  $24.12\pm 0.28$ \\ 
2005 Dec 26.2013 & 5.1240 & $i'$ &  $<24.50$  \\
2005 Dec 26.2150 & 5.1377 & $z'$ &  $<24.13$  \\
2005 Dec 26.2280 & 5.1518 & $r'$ &  $24.81\pm 0.21$ \\
2005 Dec 27.2030 & 6.1257 & $i'$ &  \nodata  \\
2005 Dec 27.2194 & 6.1421 & $z'$ &  \nodata  \\
2005 Dec 27.2358 & 6.1577 & $r'$ &  $<24.74$  \\
2005 Dec 29.2046 & 8.1273 & $r'$ &  $<24.92$ \\
2005 Dec 30.2042 & 9.1269 & $r'$ &  \nodata \\
\enddata
\tablenotetext{a}{Photometry of residual images (see \S\ref{sec:optical}).  We have assumed the source flux to be negligible in the final (template) epoch for each filter.  Errors are given as $1\sigma$ and limits are given as $5\sigma$.  These measurements are not corrected for Galactic extinction of $A_{r'}=0.19$, $A_{i'}=0.14$ and $A_{z'}=0.12$ mag \citep{sfd98}.}
\label{tab:gemini}
\end{deluxetable}

\clearpage

\begin{deluxetable}{lrrrc}
\tablecaption{Spectroscopic lines for GRB\,051221a}
\tablewidth{0pt} \tablehead{
\colhead{Line} & \colhead{$\lambda_{\rm rest}$} & \colhead{$\lambda_{\rm obs}$\tablenotemark{a}} & \colhead{Redshift} & \colhead{Flux\tablenotemark{b}} \\
\colhead{} & \colhead{(\AA)} & \colhead{(\AA)} & \colhead{} & \colhead{($\times 10^{-17}\rm erg~cm^{-2}~s^{-1}$)}
}
\startdata
$\rm [OII]$ & 3728.48   &      5765.68    &   0.5464 &  10.3  \\
H$\gamma$   & 4341.69   &      6714.53    &   0.5465 &  1.1   \\
H$\beta$    & 4862.69   &      7519.62    &   0.5464 &  3.9  \\
$\rm [OIII]$  & 4960.24   &      7671.23    &     0.5465    &      1.4  \\
$\rm [OIII]$  & 5008.24   &      7743.78    &     0.5462    &      4.5  \\
CaII K  & 3934.77   &      6082.33    &     0.5458    &      0.15\tablenotemark{c} \\      
CaII H  & 3969.59   &      6136.58    &     0.5459    &      0.14\tablenotemark{c} \\
\enddata
\tablenotetext{a}{Observed wavelengths have been corrected to vacuum.}
\tablenotetext{b}{Flux values have not been corrected for Galactic extinction}.
\tablenotetext{c}{Fluxes are measured at the line bottom following the prescription of \citet{ros85}.}
\label{tab:lines}
\end{deluxetable}

\clearpage

\begin{deluxetable}{lrrr}
\tablecaption{{\it Swift}/XRT Observations of GRB\,051221a}
\tablewidth{0pt} \tablehead{
\colhead{$\Delta t$} & \colhead{sample time} & \colhead{$F_X$ (0.3-10 keV)} & \colhead{$\sigma$} \\
\colhead{(days\tablenotemark{a})} & \colhead{(days)} & \colhead{($\rm erg~cm^{-2}~s^{-1}$)} & \colhead{($\rm erg~cm^{-2}~s^{-1}$)}
}
\startdata
$1.12\times 10^{-3}$ & $1.16\times 10^{-4}$ & $2.02\times 10^{-10}$ & $3.82\times 10^{-11}$ \\
$1.91\times 10^{-3}$ & $1.48\times 10^{-3}$ & $5.83\times 10^{-11}$ & $1.76\times 10^{-11}$ \\
$3.40\times 10^{-3}$ & $1.48\times 10^{-3}$ & $3.88\times 10^{-11}$ & $5.54\times 10^{-12}$ \\
$4.88\times 10^{-3}$ & $1.48\times 10^{-3}$ & $2.51\times 10^{-11}$ & $3.33\times 10^{-12}$ \\
$6.37\times 10^{-3}$ & $1.48\times 10^{-3}$ & $2.08\times 10^{-11}$ & $2.91\times 10^{-12}$ \\
$7.85\times 10^{-3}$ & $1.48\times 10^{-3}$ & $1.63\times 10^{-11}$ & $2.57\times 10^{-12}$ \\
$9.34\times 10^{-3}$ & $1.48\times 10^{-3}$ & $1.38\times 10^{-11}$ & $2.37\times 10^{-12}$ \\
$1.08\times 10^{-2}$ & $1.48\times 10^{-3}$ & $1.02\times 10^{-11}$ & $2.03\times 10^{-12}$ \\
$1.60\times 10^{-2}$ & $5.93\times 10^{-3}$ & $7.33\times 10^{-12}$ & $8.63\times 10^{-13}$ \\
$2.19\times 10^{-2}$ & $5.93\times 10^{-3}$ & $4.98\times 10^{-12}$ & $7.19\times 10^{-13}$ \\
$2.78\times 10^{-2}$ & $5.93\times 10^{-3}$ & $3.68\times 10^{-12}$ & $6.96\times 10^{-13}$ \\
$6.19\times 10^{-2}$ & $4.05\times 10^{-2}$ & $1.97\times 10^{-12}$ & $2.78\times 10^{-13}$ \\ 
0.102 & $4.05\times 10^{-2}$ & $2.06\times 10^{-12}$ & $2.97\times 10^{-13}$ \\
0.143 & $4.05\times 10^{-2}$ & $2.16\times 10^{-12}$ & $2.21\times 10^{-13}$ \\
0.183 & $4.05\times 10^{-2}$ & $1.70\times 10^{-12}$ & $6.43\times 10^{-13}$ \\
0.223 & $4.05\times 10^{-2}$ & $1.33\times 10^{-12}$ & $1.75\times 10^{-13}$ \\
0.264 & $4.05\times 10^{-2}$ & $9.55\times 10^{-13}$ & $1.95\times 10^{-13}$ \\
0.305 & $4.05\times 10^{-2}$ & $9.71\times 10^{-13}$ & $2.17\times 10^{-13}$ \\
0.345 & $4.05\times 10^{-2}$ & $7.46\times 10^{-13}$ & $1.28\times 10^{-13}$ \\
0.427 & $4.05\times 10^{-2}$ & $1.14\times 10^{-12}$ & $1.70\times 10^{-13}$ \\
0.522 & 0.116 & $6.71\times 10^{-13}$ & $9.69\times 10^{-14}$ \\
0.638 & 0.116 & $5.59\times 10^{-13}$ & $8.63\times 10^{-14}$ \\
1.18 & 0.480 & $1.41\times 10^{-13}$ & $2.29\times 10^{-14}$ \\
1.66 & 0.480 & $1.55\times 10^{-13}$ & $2.45\times 10^{-14}$ \\
2.27 & 0.656 & $1.09\times 10^{-13}$ & $1.65\times 10^{-14}$ \\
2.93 & 0.656 & $1.15\times 10^{-13}$ & $1.67\times 10^{-14}$ \\
3.58 & 0.656 & $7.64\times 10^{-14}$ & $1.40\times 10^{-14}$ \\
4.45 & 0.990 & $4.45\times 10^{-14}$ & $8.57\times 10^{-15}$ \\
5.44 & 0.990 & $4.88\times 10^{-14}$ & $8.92\times 10^{-15}$ \\
6.42 & 0.990 & $3.72\times 10^{-14}$ & $7.93\times 10^{-15}$ \\
7.44 & 0.957 & $3.08\times 10^{-14}$ & $7.94\times 10^{-15}$ \\
8.45 & 0.953 & $4.58\times 10^{-14}$ & $8.82\times 10^{-15}$ \\
11.33\tablenotemark{b} & 4.72 & $1.99\times 10^{-14}$ & $3.48\times 10^{-15}$ \\
\enddata
\label{tab:xrt}
\tablenotetext{a}{$\Delta t$ values give at mid-exposure time.}
\tablenotetext{b}{Co-addition of multiple observations.}
\end{deluxetable}

\clearpage

\begin{deluxetable}{lrrr}
\tablecaption{Radio Observations of GRB\,051221a}
\tablewidth{0pt} \tablehead{
\colhead{Date Obs} & \colhead{$\Delta t$} & \colhead{$F_{\nu,8.46~\rm GHz}$} & \colhead{$\sigma_{\rm 8.46~\rm GHz}$\tablenotemark{a}} \\
\colhead{(UT)} & \colhead{(days)} & \colhead{($\mu$Jy)} & \colhead{($\mu$Jy)}
}
\startdata
2005 Dec 21.99 & 0.91 & 155 & $\pm 30$ \\
2005 Dec 23.02 & 1.94 & \nodata & $\pm 24$ \\
2005 Dec 24.83 & 3.75 & \nodata & $\pm 32$ \\
2005 Dec 27.96 & 6.88 & \nodata & $\pm 28$ \\
2006 Jan 14.01 & 23.93 & \nodata & $\pm 16$ \\
\enddata
\tablenotetext{a}{All errors are given as $1\sigma$ (rms).}  
\label{tab:vla}
\end{deluxetable}

\clearpage

\begin{deluxetable}{lc}
\tablecaption{Physical Parameters for GRB\,051221a}
\tablewidth{0pt} \tablehead{
\colhead{Parameter} & \colhead{Value}
}
\startdata
$E_{\gamma,\rm iso}$ & $2.4^{+0.1}_{-1.3}\times 10^{51}$ erg \\
$E_{K,\rm iso}$ & $(1.1 - 1.6)\times 10^{51}$ erg \\
$n$ & $(0.5 - 2.4)\times 10^{-3}~~\rm cm^{-3}$ \\ 
$\epsilon_e$\tablenotemark{a} & (0.24 - 1/3) \\ 
$\epsilon_B$\tablenotemark{a} & (0.12 - 1/3) \\
$\theta_j$ & $ (5.7 - 7.3)^{\circ}$ \\
$f_b$ & $ (0.005 - 0.008)$ \\
$E_{\gamma}$ & $(1.2 - 1.9)\times 10^{49}$ erg \\
$E_{K}$ & $(7.8 - 8.9)\times 10^{48}$ erg \\ 
\enddata
\tablenotetext{a}{Values are constrained to be $< 1/3$.}  
\label{tab:params}
\end{deluxetable}

\clearpage

\end{document}